\documentclass[aps,showpacs,pra,floatfix,superscriptaddress,twocolumn,10pt]{revtex4-1}
\usepackage{graphicx}
\usepackage{amsmath}
\usepackage{txfonts}
\usepackage{bm} % load after all math to give access to bold math

\begin{document}
\title{Destruction of Long-range Order by Quenching the Hopping Range in One Dimension}
\author{Masaki Tezuka} \email{tezuka@scphys.kyoto-u.ac.jp}
\affiliation{Department of Physics, Kyoto University, Kitashirakawa, Sakyo-ku, Kyoto 606-8502, Japan}
\author{Antonio M. Garc\'{\i}a-Garc\'{\i}a}
\affiliation{Cavendish Laboratory, University of Cambridge, JJ Thomson Avenue, Cambridge, CB3 0HE, UK}
\affiliation{CFIF, Instituto Superior T{\'e}cnico,
Universidade de Lisboa, Av. Rovisco Pais, 1049-001 Lisboa, Portugal}
\author{Miguel A. Cazalilla}
\affiliation{Department of Physics, National Tsing Hua University,
and National Center for Theoretical Sciences (NCTS), Hsinchu City, Taiwan}
\begin{abstract}
We study the dynamics in a one dimensional hard-core Bose gas with power-law hopping after an abrupt reduction of the hopping range using the time-dependent density-matrix renormalization group (t-DMRG) and bosonization techniques. In particular, we focus on the destruction of the Bose-Einstein condensate (BEC), which is present in the initial state in the thermodynamic limit. We argue that this type of quench is akin to a sudden reduction in the effective dimensionality $d$ of the system (from $d > 1$ to $d = 1$). We identify two regimes in the evolution of the BEC fraction. For short times the decay of the BEC fraction is Gaussian while for intermediate to long times, it is well described by a stretched exponential with an exponent that depends on the initial effective dimensionality of the system.  These results are potentially relevant for cold trapped-ion experiments which can simulate an equivalent of hard-core bosons, i.e. spins, with tunable long-range interactions. 
\end{abstract}
\date{\today}
\maketitle
Many physical phenomena occur under non-equilibrium conditions. Small deviations from equilibrium are well understood in the framework of linear response theory. However the time evolution of non-equilibrium states has resisted so far a comprehensive understanding despite its potential to open new avenues of research in several fields including solid condensed matter and cold atom physics as well as cosmology.

A number of recent advances are changing this situation rapidly. 
In condensed matter, experiments \cite{far1,far2,far3,far4} that combine femtosecond extreme ultraviolet pulses with very sensitive time and angle resolved photo-emission spectroscopy are able to probe the far from equilibrium dynamics of many-electron systems. Relevant examples include the time evolution of the order parameter \cite{far2} characterizing superconductivity or charge density wave (CDW) order \cite{far3,far4} or the discovery of transient superconductivity in cuprates \cite{far1}. 

In the context of ultracold atom physics, it has been observed experimentally \cite{kinoshita} that the momentum distribution of a quasi-one-dimensional (1D) Bose gas that is initially prepared in a far-from equilibrium state does not exhibit thermalization. More recently, the loss of phase coherence for sufficiently long times, after a sudden decoupling of two phase coherent 1D Bose condensates was investigated experimentally in \cite{mayer1} and theoretically in \cite{altman,demler1}. On the theory side, it has been shown  that integrable models in general fail to thermalize to the standard Gibbs ensemble but relax instead to a generalized Gibbs ensemble (GGE) \cite{rigol,cazalilla2012}. This was shown by Rigol and coworkers in Ref.~\cite{rigol} using numerical simulations for the XX model and subsequently analytically shown for a quantum quench in the Luttinger model by one of us~\cite{miguel}. Furthermore, quantum quenches from a non-critical to a critical state were studied by Calabrese and Cardy~\cite{cardy} using a clever mapping to a boundary conformal field theory. They found that the order parameter would decay exponentially after the quench. 

Indeed, a quench is a convenient procedure to study non-equilibrium dynamics by making a sudden change of either a coupling constant in the Hamiltonian or an external parameter such as temperature or magnetic field. In some cases, the sudden removal of an external field or a change in a coupling constant can lead to destruction of long-range order in the initial state.  In this article, we report an example of this, in which the parameter being quenched is the range of the hopping amplitude (from long to short range) in a one-dimensional (1D) interacting Bose system. As we argue below, this type of quench has some similarities with a sudden change in the effective dimensionality of the system from $d > 1$ to $d = 1$.   We show using both numerical (i.e. DMRG) and analytical (i.e. bosonization) techniques that this type of quench leads to an interesting non-equilibrium regime in the time evolution of a Bose gas at zero temperature.

In order to mimic a higher spatial (i.e. $d > 1$) dimensionality while retaining both the numerical and analytical convenience  of one dimension, we assume that the system is a 1D Bose gas with long-range power-law hopping in the initial state. By now there is solid evidence from studies in different physical contexts \cite{mirlin,chaos,levitov,hublong} that long-range hopping is an effective way of mimicking higher dimensional effects in one dimensional systems. It is therefore reasonable to expect that power-law hopping in our model will also resemble an effective, not necessarily integer, dimensionality $d>1$, which yields a true BEC~\cite{uslobos} in the ground state for sufficiently slow decay (effective dimensionality). As we explain below, such dimensionality quench can be realized in an optical lattice of coupled 1D gas tubes~\cite{cazalilla2006}. An additional motivation for this type of quenches stems from the recent realization of spin-chains in trapped ion systems with highly-controllable long-range interactions~\cite{spinc}. In this regard, we recall that in spins and bosons are related by an exact mapping~\cite{cazalilla2011r}. 

In our calculations, to be described below, we have observed two different regimes in the evolution of the condensate fraction.  
For short times, the depletion of BEC is consistent with a Gaussian decay, whereas
at intermediate to long times, it appears to cross over to a stretched exponential behavior. These two limiting behaviors are also obtained from a bosonization~\cite{cazalilla2011r} approach. Consistent with this description, which treats phonons (i.e. the low energy excitations of the post-quench Hamiltonian) as propagating ballistically, we believe this regime corresponds to a prethermalized regime.

The rest of the article is organized as follows: In Sect.~\ref{sec:model} we describe the model of that we have studied. Details about the quench protocol and the way the BEC fraction is measured using DMRG are given in Sect.~\ref{nume}. The details of the analytical approach are presented in Sect.~\ref{sec:analytic}. The conclusions of this work is given in Sect.~\ref{sec:concl}. In Appendix~\ref{sec:convergence} we give further detail on the numerical convergence of the t-DMRG calculation.

\section{Model and quench} \label{sec:model}
The initial state before the quench is described by the ground state of the following Hamiltonian of $N$ hard-core bosons hopping on a 1D lattice:
\begin{equation}
\hat{H}_\mathrm{i} = - \sum_{m,r\geq1}^{L} J_{r} \left[ \hat{b}^{\dag}_{m+r} \hat{b}_m +{\mathrm{H.c.}} \right]
+ V\sum_{m} \hat{n}^{\dag}_{m}\hat{n}_{m+1}, \label{eq:ham1}
\end{equation}
in which $V$ is the strength of the nearest-neighbor density-density interaction, $\hat{n}_m \equiv \hat{b}^{\dag}_{m}\hat{b}_m$ and
\begin{equation}
J_{r} = J\left[ \delta_{r,1} + \frac{f}{r^{\kappa}} (1-\delta_{r,1})\right] \label{eq:hopr}
\end{equation}
is the hopping amplitude. In Ref.~\cite{uslobos}, two of the present authors studied a 1D fermionic model with attractive interactions and power-law hopping characterized by an exponent $\alpha = \kappa/2$, which in the limit of strong attractive interactions exhibits similar physics to the Hamiltonian Eq. \eqref{eq:ham1} 
in the low-energy sector. In Ref.~\cite{uslobos}, long-range phase coherence was found for $\kappa \sim 2 \alpha \lesssim 3$. A detailed comparison of the boundaries between the phase exhibiting 
off-diagonal long range order and the disordered phase
 in this model and the model in Ref. 17
will be presented elsewhere. However, for 
the purpose of studying quantum quenches from
the ordered ground state, it is sufficient to choose
$\kappa$ sufficiently small and $V$ negative. Indeed,
it can be seen in Fig.~\ref{fig1} that
the off-diagonal long range order, which reflects into
the condensate fraction, is enhanced for small $\kappa$.
The (initial) condensate fraction is also made larger
by choosing a negative value of the nearest neighbor
interaction, $V$. Physically, this is because a negative $V$ weakens the phase disordering effect of the hard-core repulsion, as it is also well known for models with short
range hopping~\cite{cazalilla2011r}.  In addition, empirically we also found that larger starting condensate fraction also improves the numerical convergence in DMRG. 
   
We note that long-range interactions are of interest in a broad variety of problems: spin-chains with long-range interactions \cite{spinc}, non-interacting weakly disordered systems \cite{mirlin}, quantum chaos \cite{chaos}, systems controlled by dipolar interactions \cite{levitov} and certain types of Mott metal-insulator transitions \cite{hublong}.  In particular, it is also believed that long range interactions provide an effective way of accounting for higher dimensional effects in one dimensional systems. This is because long-range interactions provide a way around the Mermin-Wagner theorem, which forbids the existence of spontaneous breaking of continuum symmetries long range in 1D quantum systems. Thus, as mentioned above, the model \eqref{eq:ham1} exhibits long range phase coherence~\cite{uslobos} in its ground state.

Below, we consider a quantum quench at $t = 0$, in which the long range tail (i.e. the term proportional to $f$ in Eq.~\eqref{eq:hopr}) is suddenly turned off.
In other words, we shall assume that the system is prepared in the ground state of Eq.~\eqref{eq:ham1} with a $\kappa \leq 2$ such that a bona-fide Bose-Einstein condensate exists.
Following the quench, the time evolution of the Hamiltonian is described by the hard-core Bose-Hubbard model,

\begin{align}
\label{eq:ham2}
\hat{H}(t > 0) = \hat{H}_{\mathrm{f}}
= - J \sum_{m} \left[ \hat{b}^{\dag}_m \hat{b}_{m+1} + \hat{b}^{\dag}_{m+1} \hat{b}_m \right]  + V\sum_{m} \hat{n}^{\dag}_{m}\hat{n}_{m+1}.
\end{align}

Thus, we expect that, since the ground state of $\hat{H}_\mathrm{f}$ lacks long-range phase coherence (i.e. is not a BEC), but the latter is present in the initial state, the BEC will be destroyed by the time evolution under $\hat{H}_\mathrm{f}$. The question that we address here is how this destruction takes place. 

However, before describing the results of the DMRG simulation, it is worth dwelling on the role played by the parameter $\kappa$ as an effective dimensionality. To this end, we can draw an analogy with a quantum quench in an array of 1D bosons. Let us assume that each 1D system is described by a copy (labeled by a new index $i$) of $\hat{H}_\mathrm{f}$ and is initially coupled to $z$ nearest neighbors by means of the following Josephson tunneling term:
\begin{equation}
\hat{H}_\mathrm{J} = - J_{\perp} \sum_{\langle i, j
\rangle}\sum_{m}  \left[ \hat{b}^\dag_{m}(i) \hat{b}_{m}(j) + \mathrm{H.c.} \right],
\end{equation}
where $\sum_{\langle i, j\rangle}$ indicates sum over nearest neighbors and  $J_{\perp} \ll J_1$.  Let $z$ approach infinity while $z J_{\perp}$ remains constant. In this limit, which effectively amounts to infinite dimensions, the mean-field approximation becomes exact and, the Josephson coupling term becomes (up to a constant) $H^{\mathrm{MF}}_\mathrm{J} = \sum_{i} \hat{H}_\mathrm{J}(i)$, in which $\hat{H}_\mathrm{J}(i) =  -z J_{\perp}  b_0  \sum_{m} \left[\hat{b}^{\dag}_m + \hat{b}_m\right]$, with $b_0 = \langle \hat{b}_m \rangle$. Using bosonization~\cite{cazalilla2011r}, it is possible to map the mean-field Hamiltonian in the continuum limit to a sine-Gordon model~\cite{cazalilla2006,cazalilla2011r}. In this setup, a quench where the Josephson coupling $J_{\perp}$ is suddenly turned off corresponds to  suddenly turning off the sine-Gordon term in the continuum model, a problem that has attracted much interest in recent times~\cite{cardy,iucci2010,mitra2012}.
Since the quench in the sine-Gordon model is from a gapped to a critical state, we can rely on the boundary conformal field theory results of Calabrese and Cardy~\cite{cardy}, which predict that the order parameter will decay as:
\begin{equation}
\langle \hat{b}_m(t) \rangle \sim e^{-t/\tau}
\label{eq:cardy}
\end{equation}
at asymptotically long times after the quench. In the above expression, $\tau$ is related to the size of the gap in the initial non-critical state (which corresponds to the healing length of the initial condensate~\cite{cazalilla2011r}). This behavior bears some resemblance to the result derived below using bosonization for the model with long-range hopping in Eq.~\eqref{eq:ham1}, which is a stretched exponential for a generic value of $\kappa$ (cf.~Eq.~\eqref{sube}).
Indeed, tentatively we may assume that the limit $z\to +\infty$ corresponding to infinite dimensions is akin to setting $\kappa = 1$. Thus, in general, we can regard the role of the parameter $\kappa$ as controlling the effective dimensionality of the system. This allows for a study of how long-range phase coherence is suppressed by fluctuations after an abrupt change in the dimension of the system.

However, the theory of Ref.~\cite{cardy} presumes that (boundary) conformal field theory is an exact description of the critical state. This ignores the existence of an infinite set of irrelevant operators that are responsible for the breaking of the infinite set of conservation laws that characterize the conformal field theory. Therefore, Eq.~\eqref{eq:cardy} is expected to hold at intermediate times, only. For longer times, the system will eventually transition to a different state. However, since $\hat{H}_\mathrm{f}$ is an integrable model~\cite{cazalilla2011r}, the nature of such a state is not clear at the moment. 

\section{Numerical results} \label{nume}
\begin{figure}
\includegraphics[]{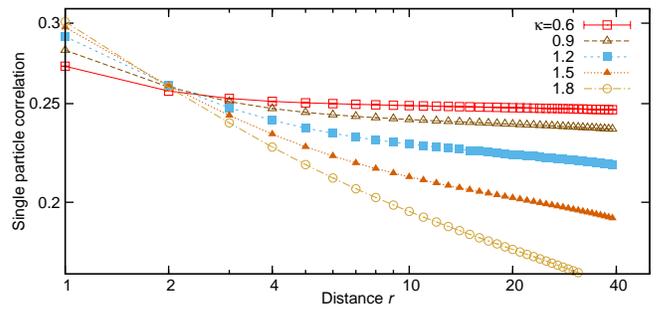}
\caption{(Color online) Single particle correlation function  $\langle \hat{b}(j+r)^\dag \hat{b}(j) \rangle_j$ for the Hamiltonian Eq.(\eqref{eq:ham1}) for $L=80$, $N=40$, $V=-1.2$, $f=0.1$ and different values of $\kappa$ computed using DMRG. In agreement with the theoretical expectation, phase coherence is observed up to $\kappa \sim 2$. For larger $\kappa$ the crossover length to observe it is larger than the maximum size accessible.
In order to avoid these unwanted size effects the initial state, previous to the quench, will be always characterized by $\kappa \leq 2$. 
}
\label{fig1}
\end{figure}

We simulate the quench dynamics using the time-dependent DMRG (t-DMRG) \cite{dmrg}.
Although the pre-quench Hamiltonian contains terms involving power-law hopping between distant lattice sites, which make DMRG calculations less efficient, the post-quench Hamiltonian contains only hopping to nearest-neighbors. Therefore, the efficient time evolution algorithm based on the Suzuki-Trotter breakup \cite{dmrg} can be employed.  First of all, we prepare the initial state and check that, to a good approximation, phase coherence holds.
In Fig.~\ref{fig1}, we have plotted the spatial average of the single particle correlation function $\langle\langle \hat{b}(j+r)^\dag \hat{b}(j) \rangle \rangle_j$  against the distance $r$ for different values of $\kappa$ obtained by t-DMRG. 
As expected, long-range phase coherence, which manifest itself in a small variation of the single-particle correlation function, is present for sufficiently small $\kappa$. As described above, the initial state is therefore akin to a Bose gas with short-range hopping and an effective higher dimensionality.  

The condensate fraction $q(t)$ is given by the largest eigenvalue of the
two-particle correlation $M_{i,j}(t) \equiv \langle
\Psi(t)|\hat{b}^\dag_i \hat{b}_j|
\Psi(t) \rangle$ divided by the number of bosons $N$ ($N=40$ and $L=80$ in all the simulation reported below). We set
$f = 0.1$ and $V = -1.2$ where $J_r = fr^{-\kappa}J$ for $r\geq 2$ in \eqref{eq:ham1}. We have observed that the numerical error accumulates faster for smaller values of $\kappa$.
For $\kappa = 1$, up to time $t\sim 4/J$, the results with DMRG block dimensions $M = 600$ and $700$
are almost identical, indicating a good convergence for all the values
of $\kappa$. The unit of time is $1/J$  (we take $\hbar = 1$).  We have also checked that conservation of energy holds in this time interval though small violations $< 5\%$ are already observed for small $\kappa \sim 1$ and $t\sim 4/J$. 
The time step is $\Delta t = 0.025/J$  in all simulations. 
We have also confirmed that the results using $\Delta t = 0.04/J$
and $M = 500$ also coincide
with the data shown up to $t \sim 4.5 / J$.
Generally speaking, t-DMRG is reliable only for times for which $q(t) \gg 1/N$, which also limits the time interval that can be explored numerically. 
 
\begin{figure}
\includegraphics{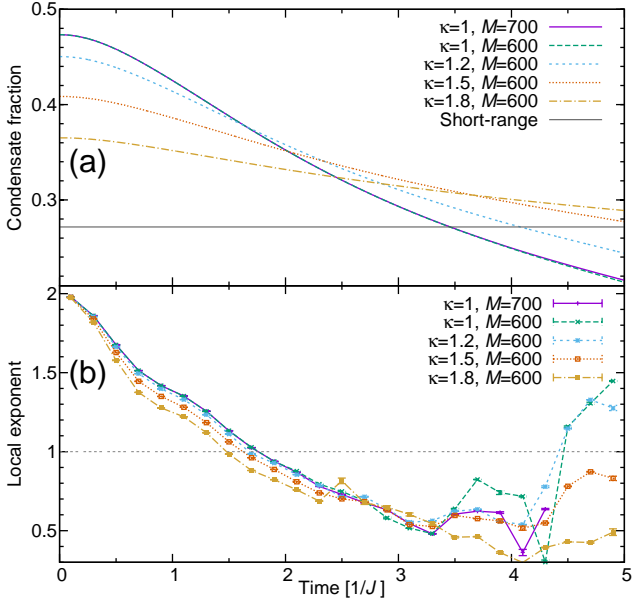}
\caption{(Color online)
a) The condensate fraction $q(t)$ for different values of $\kappa$ after a quench of the Hamiltonian Eq.~\eqref{eq:ham1} with $L=80$ and $N=40$. The nearest neighbor interaction between the hard-core bosons is $V=-1.2$, and the pre-quench power-law hopping coefficient is $f=0.1$. After the quench the hopping is restricted to nearest neighbors which corresponds to the $\kappa \to \infty$ limit. b) Results of the non-linear fitting of the t-DMRG results against the function $q_\mathrm{fit}(t) = q_0 \exp{(-\gamma t^c)}$. The fitting was carried out in small time intervals of length $0.2/J$ centered at times $0.1/J, 0.3/J, ..., 4.9/J$. The fitted value of the time dependent exponent $c(t)$ depicted in the figure plotted against the center of each segment. For short ($\ll 1/J$) times $c(t)$ is almost two, as expected in Eq.~\eqref{gauss}.
For intermediate times $\approx 3/J$ we find qualitative agreement with the analytical prediction Eq.~\eqref{sube},
while for longer times, especially for small $\kappa$ an expected deviation of $c(t)$ is observed. See text for more details.} 
\label{fig2}
\end{figure}

With these technical limitations in mind, we plot in
Fig.~\ref{fig2} the time evolution of the condensate fraction $q(t)$ for
$V=-1.2$ and $1 \leq \kappa \leq 1.8$.
The decay is faster as $\kappa$ increases.
Qualitatively this is understood as follows: A larger $\kappa$ means that the system is closer to the post-quench Hamiltonian therefore the amount of energy injected into the system by the quench is smaller. 

\section{Analytical Results}\label{sec:analytic}
In this section, we study the bosonized form of Eq.~\eqref{eq:ham1}. To this end, we write $\hat{b}_{m} = \sqrt{\rho + \delta n_m} e^{i\theta(x_m)} \simeq \sqrt{\rho} e^{i \theta(x_m)}$ and $\hat{n}_m = \rho + \frac{1}{\pi} \partial_x \phi(x_m) + \cdots$, in which $x_m = m$, $\theta(x)$ (the phase field) and  $\partial_x \phi(x)$ (the density field) are canonically conjugate to each other~\cite{cazalilla2011r}; the mean lattice occupation is $\rho = \frac{N}{L}$ ($=\frac{1}{2}$ in the numerical simulation of Sect.~\ref{nume}). In this section, we consider a translationally invariant system with $L\to +\infty$ and  $\rho = \mathrm{constant}$. In terms of $\phi(x)$ and $\theta(x)$, the low-energy effective Hamiltonian reads:
\begin{align}
\hat{H}_\mathrm{i}^\mathrm{eff} &= \frac{v}{2\pi}\int dx \left[ K \left(\partial_x \theta(x) \right)^2 + K^{-1} \left( \partial_x \phi(x) \right)^2 \right]  \nonumber  \\ &- \rho \int_{|x-x^{\prime}| > a} 
dx dx^{\prime} \, t(x-x') \cos
\left[ \theta(x)  - \theta(x^{\prime}) \right], \label{eq:ham3}
\end{align}
where $v$ is the sound velocity, $K$ the Luttinger parameter, and $t(x-x') \simeq f/|x-x'|^{\kappa}$ ($a\approx 1$ is a short distance cut-off). The last term is a non-linear function of the phase field, $\theta(x)$, which makes the task of computing the ground state by elementary methods impossible. However, a reasonable approximation to the ground state of \eqref{eq:ham3} when the long-range boson hopping is a relevant (in the renormalization-group sense) perturbation can be obtained using the self-consistent harmonic approximation (SCHA)~\cite{uslobos}, which amounts to replacing the last term in the right hand-side of \eqref{eq:ham3} by 
\begin{equation}
\frac{1}{2}\int_{|x-x^{\prime}| > a} dx dx^{\prime} T(x-x') \left[ \theta(x) - \theta(x^{\prime}) \right]^2,
\end{equation}
where the function $T(x-x') \sim 1/|x-x'|^{\kappa}$ must be determined variationally~\cite{uslobos}. In order to diagonalize the quadratic Hamiltonian resulting from the SCHA we expand the phase field as $\theta(x) = \theta_0 + \frac{n \pi x}{L} + \Theta(x) + \Theta^{\dag}(x)$ with $n$ the winding number, $\Theta(x) =  \frac{1}{2}\sum_{q\neq 0} \left( \frac{2\pi}{K |q| L}\right)^{1/2} \mathrm{sgn}(q) \: e^{i q x}\:  \hat{\tilde b}_q$ and $\left[ \hat{\tilde b}_{q}, \hat{\tilde b}^{\dag}_{q^{\prime}} \right] = \delta_{q,q^{\prime}}$.
The resulting Hamiltonian can be diagonalized by a new set of boson eigenmodes described by $(\hat{a}_q, \hat{a}^{\dag}_q)$, and for $L\to +\infty$, it reads:
\begin{align}
\hat{H}_\mathrm{i}^\mathrm{eff} \simeq E_0 + \sum_{q\neq 0} \omega(q) \hat{a}^{\dag}_q \hat{a}_q
\end{align}
where the constant $E_0$ is the ground state energy and $\omega(q) = \sqrt{v^2 q^2 + v\pi \rho T(q)/K}$. For $q \ll a^{-1}$, $\omega(q) \sim T^{1/2}(q) \sim |q|^{\frac{\kappa -1}{2}}$ since $T(q)\sim |q|^{\kappa - 1}$ for $\kappa < 3$.

In the following, we focus on the time evolution at times $t > 0$ of the phase field $\theta(x)$ under the Hamiltonian $\hat{H}_\mathrm{f}^\mathrm{eff} = \sum_{q\neq 0} v |q| \hat{\tilde b}^{\dag}_q \hat{\tilde b}_q$ corresponding to a 1D Bose gas with short-range hopping~\cite{cazalilla2011r}:
\begin{align}
\theta(x,t) &=  e^{i \hat{H}_\mathrm{f}^\mathrm{eff} t} \theta(x) e^{- i \hat{H}_\mathrm{f}^\mathrm{eff} t}  \\
 &=   \frac{1}{2}\sum_{q\neq 0}\left( \frac{2\pi}{K |q| L}\right)^{1/2} 
\mathrm{sgn}(q) e^{i q x}\left[ e^{-i v|q| t} \hat{\tilde b}_q - e^{+i v |q| t}\hat{\tilde b}^{\dag}_{-q} \right] ,
\end{align}
We emphasize that, within SCHA, $(\hat{a}_q, \hat{a}^{\dag}_q)$ and $(\hat{\tilde b}_q, \hat{\tilde b}^{\dag}_q)$ are the eigenmodes of two quadratic Hamiltonians and therefore they can be related by a  canonical transformation~\cite{cazalilla2012}:
\begin{equation}
\hat{\tilde b}_q  = u_q \hat{a}_q + v_q \hat{a}^{\dag}_{-q}, \quad \hat{\tilde b}^{\dag}_{-q} =  v_q \hat{a}_{q} + u_{q} \hat{a}^{\dag}_{-q}.
\end{equation}
in which $u_q = \cosh \theta_q$ and $v_q = \sinh \theta_q$, with $\theta_q$ satisfying
\begin{equation}
\tanh 2\theta_q = \frac{2 u_q v_q}{u^2_q + v^2_q} = \frac{\pi \rho T(q)/(|q| K)}{v|q|  + \frac{\pi\rho T(q)}{|q|K}}.
\end{equation}
 
In order to compute the evolution of the BEC order parameter, $\langle \hat{b}_m(t) \rangle \simeq \rho \:  \langle e^{i\theta(m,t)} \rangle  = \rho \:  e^{-\frac{1}{2} \langle \theta^2(0,t)\rangle}$ we need to obtain  $\langle \theta^2(0,t)\rangle$, where $\langle \ldots \rangle$ stands for the average over the ground state of the SCHA to the initial Hamiltonian,  Eq.~\eqref{eq:ham1}. Asymptotically, at long times we find:
\begin{align}
\langle \theta^2(0,t) \rangle  - \langle \theta^2(0,0) \rangle  
 &\sim \int^{\Lambda}_{1/(vt)} dq \, \frac{T^{1/2}(q)}{q^2}  \nonumber \\
 & \sim  \int^{\Lambda}_{1/(v t)} a_0 dq \: |a_0 q|^{\frac{\kappa-5}{2}}  \sim \frac{(v t/a_0)^{\frac{3-\kappa}{2}}}{ \frac{\kappa-3}{2}}.
 \label{eq:theta0}
\end{align}
where $a_0 \simeq \Lambda^{-1}$ is a short distance cut-off, of the order of the lattice parameter. 
Hence, the condensate fraction decays according to a stretched exponential law:
\begin{equation}
\langle \hat{b}_m(t) \rangle =  A_0 e^{- \gamma t^{\frac{3-\kappa}{2}}}
\label{sube},
\end{equation}
where $\gamma$ and $A_0$ are positive constants that depend on the microscopic details of the initial and final Hamiltonians. 

In the opposite limit of short times after the quench, the decay is a Gaussian:  
\begin{equation}
\langle e^{i \theta(x,t)} \rangle \sim q_0 e^{- \beta t^2} \approx q_0 (1 - \beta t^2) ,
\label{gauss}
\end{equation}
where the dependence on $\kappa$ appears only through $A_1 > 0$ and $\beta > 0$.

\section{Discussion}

In Fig.~\ref{fig2}b we compare t-DMRG results and the analytical predictions described above. We recall that, for short times, we expect $q(t)=q_0 {\mathrm e}^{-\beta t^2}$ with $q_0$ the initial condensate fraction.
%As it was mentioned previously
The parameter $\beta$ controls the typical time scale for which a mean field approach holds.
For intermediate times the theoretical prediction is $q(t)\sim {\mathrm e}^{- \gamma t^{\frac{3-\kappa}{2}}}$. 
Since the analytical techniques do not allow an accurate estimation of these parameters we have fitted the numerical $q(t)$ in small time intervals of length $0.2/J$ 
at times $0.1/J, 0.3/J, ..., 4.9/J$. We have carried out non-linear least squares fit with a fitting function
$q_\mathrm{fit}(t) = q_0 \exp{(-\gamma t^c)}$ where the time dependent exponent $c(t)$ is obtained from the fitted value at the center of each segment.

We note that the fitting interval is limited by the loss of accuracy of the t-DMRG method at long times, which manifests itself in the energy being no longer conserved and the convergence with $m$ of $q(t)$ deteriorating rapidly.
With these limitations in mind, the results depicted in Fig.~\ref{fig2}b show a resonable good agreement with the analytical expectations for the range of $\kappa$ considered. 
For short times, the local exponent $c(t) \approx 2$, as expected from analytical approach described above, while for longer times, it decreases below this value. For $t \approx 3/J$, $c(t)$ shows a clear dependence with $\kappa$, which is consistent with the theoretical prediction that $c(t) \propto \frac{3-\kappa}{2}$ at long times.
For longer times, although the calculated condensate fraction $q(t)$ is almost converged with respect to $m$,  $c(t)$ is not presumably due to the high sensitivity of $c(t)$ to the local change of $q(t)$ via the non-linear fitting process, therefore it is not meaningful to compare the values of $c(t)$ beyond $t \approx 3/J$.

However, we note that deviations from the stretched exponential behavior of Eq.~\eqref{sube}, obtained from the bosonization approach of Sect.~\ref{sec:analytic}, are also to be expected theoretically. In fact, in a infinite system, the stretched exponential behavior shows up during the prethermalized regime~\cite{kehrein,mitra2012,nessi2014}. In such a regime, the phonons described by $\{\hat{\tilde{b}}_q, \hat{\tilde{b}}^{\dag}_q\}$, which are the low-energy elementary excitations of the post-quench Hamiltonian, $H_f$, propagate ballistically. However, at longer times, the phonons will scatter each other, which should lead to a different decay of the BEC fraction~\cite{mitra2012}. 
The calculation of this behavior is beyond the scope of this work, as it requires taking into account an infinite number of irrelevant operators that are neglected in the bosonization approach. Furthermore, we have not been able to determine whether the observed deviations from the sub-exponential behavior are due to the approximate treatment of the initial state and to the neglect of the open boundaries and finite size effects, which are inevitably present in the DMRG simulations.
 
For even longer times, we also expect the destruction of the initial BEC will be driven, not only by the phonons, but also by topological excitations known as phase slips. The latter are instantons, i.e. topological defects in imaginary time. Qualitatively, we expect the inverse Kibble-Zurek (KZ) \cite{kibble,zurek} mechanism, similar to the phenomenon studied recently in Ref.~\cite{polkov} in the context of a classical quench in a two-dimensional Bose gas.  According to the KZ scenario, after a quench, from a disordered phase, topological fluctuations are expected to survive for a long time in the final ordered phase. In the inverse KZ scenario the quench is from the ordered to disordered phase. The resulting dynamics is characterized \cite{polkov} by a supercooled phase for short times and a slow proliferation of vortices for longer times that leads to an unexpected resilience of the superfluid state. \\

\section{Conclusion}\label{sec:concl}
Using a combination of time-dependent DMRG and bosonization, we have studied the destruction of the Bose-Einstein condensate (BEC) in a one-dimensional Bose gas following a quench in the range of the hopping amplitude. 
We have argued that this is akin to a dimensionality quench, as the exponent characterizing the range of the hopping in the pre-quench Hamiltonian plays the role of an effective (non-integer) dimensionality $d > 1$.

Our main result is the identification of two distinct regimes for the destruction of Bose-Einstein condensate (BEC) after the quench in effective dimensionality quench described above. At short times, we find a decrease of the BEC fraction of the form $q(t) = q_0 e^{-\beta t^2} \approx q_0 (1 - \beta t^2)$, where $q_0$ is the initial BEC fraction and $\beta > 0$ is a constant. At longer times, this behavior crosses over to a different kind of time dependence. From a bosonization approach (see Sect.~\ref{sec:analytic}), we find a sub-exponential behavior $q(t) \sim {\mathrm e}^{-\gamma t^{\delta}}$ with $\delta = \frac{3-\kappa}{2}$, with $\gamma > 0$ and $\kappa$ the power-law hopping exponent which is directly related to the initial effective dimensionality of the condensate. This behavior appears to be consistent with the behavior of $q(t)$ obtained from the time-dependent DMRG calculations described in Sect.~\ref{nume}. We note, however, that our theoretical approach is expected to break down for sufficiently long times due to the phonon scattering and the proliferations of phase slips.  
\acknowledgments
The authors thank Ippei Danshita and Alejandro Lobos for fruitful discussions.
MT appreciates the hospitality of the TCM Group, Cavendish Laboratory, which MT visited under the support from the Kyoto University Global Frontier Project for Young Professionals ``John-Mung Program'', and the hospitality of NCTS (Taiwan) where parts of the work were done.
MT is also supported by KAKENHI (No. 26870284).
AMG was supported by EPSRC, grant No. EP/I004637/1, FCT, grant PTDC/FIS/111348/2009 and a Marie Curie International Reintegration Grant PIRG07-GA-2010-268172.
MAC is supported by NSC and a start-up grant from NTHU (Taiwan).
Part of numerical computation in this work was carried out at the Supercomputer Center, ISSP, University of Tokyo and Yukawa Institute Computer Facility, Kyoto University.
% References %%%%%%%%%%%%%%%%%%%%%%%%%%%%%%%%%%%%%%%%%%%%%%%%%%%%%%%%%%%%%%%%%%%%%

\newpage
\begin{widetext}
\end{widetext}

\appendix
\section{Convergence check of the t-DMRG simulation}\label{sec:convergence}

We discuss here our observations on the numerical simulation of the real-time evolution of a state under the post-quench Hamiltonian.
In this work, we apply the time-dependent DMRG method to study the evolution of the single-particle correlation governed by $\hat{H}_\mathrm{f}$, defined by Eq.~\eqref{eq:ham2},
after the power-law hopping terms are dropped from $\hat{H}_\mathrm{i}$, defined by Eq.~\eqref{eq:ham1}, at $t=0$.
The accuracy of the t-DMRG simulation for a given system is governed by the number of retained states per block $M$
and the simulation step in real time $\Delta t$.
The accuracy also depends on the parameter values, even if the system size is fixed, because the magnitude of the effect of the quench
on the ground state of $\hat{H}_\mathrm{i}$ depends on the exponent $\kappa$ and the prefactor $f$ of the power-law hopping terms,
and the typical timescale of the excitations in this pre-quench state to travel on the lattice depends on the nearest-neighbor interaction $V$,
which is still present in the post-quench Hamiltonian $\hat{H}_\mathrm{f}$.

\begin{figure}[h]
\includegraphics{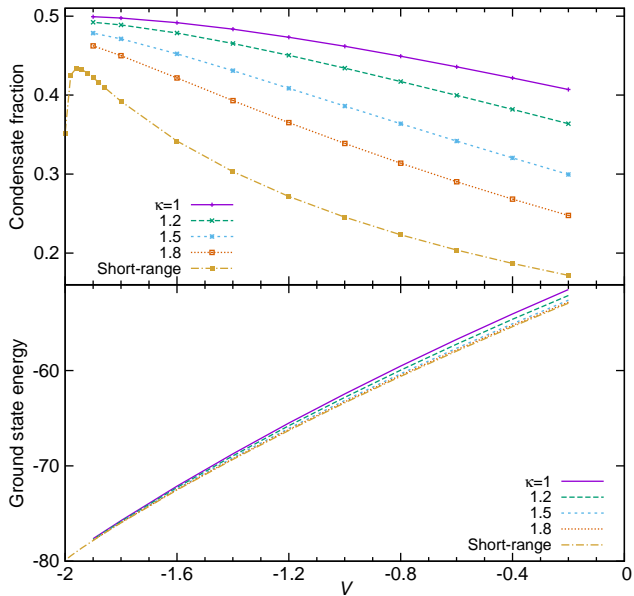}
\caption{(Color online)
Top: The calculated condensate fraction $q(t=0)$ (left axis) as a function of the nearest-neighbor interaction $V$, for $\hat{H}_\mathrm{i}$ with $f=0.1$ and $\kappa = 1, 1.2, 1.5, 1.8$ and
the post-quench Hamiltonian $\hat{H}_\mathrm{f}$ (short-range model), with $L=80$ and $N=40$.
Bottom: The energy of the pre-quench Hamiltonian ground state wave function $\vert \Psi(t=0)\rangle$ evaluated by the post-quench Hamiltonian
$\hat{H}_\mathrm{f}$, $\langle \Psi(t=0) \vert \hat{H}_\mathrm{f} \vert \Psi(t=0) \rangle$, as a function of $V$.
At least $M=500$ states have been preserved in the final finite-size system loops of the DMRG calculation.
}
\label{fig:GS1.2sr}
\end{figure}

The typical timescale of the dynamics should be governed by the inverse of the speed of sound $v$, which in turn is determined, along with the Luttinger parameter $K$, by the value of $V$.
For $V<0$, as $|V|$ is increased, $v$ is decreased. Therefore, in a system with a larger $|V|$, the condensate should decay more slowly.
The ground state of the post-quench Hamiltonian, $\hat{H}_\mathrm{f}$, at half filling has a density-wave order for $V < -2$.\cite{YangYang1966}
In the following we discuss the dependence of the dynamics on the choice of $V$ within $(-2,0)$.

The calculated condensate fraction $q(t=0)$ for the ground state $\vert \Psi(t=0)\rangle$ of $\hat{H}_\mathrm{i}$ and
the energy of this state for the post-quench Hamiltonian $\hat{H}_\mathrm{f}$, $\langle \Psi(t=0) \vert \hat{H}_\mathrm{f} \vert \Psi(t=0) \rangle$,
is plotted for various values of $\kappa$ and $f=0.1$ and for the short-range case ($\hat{H}_\mathrm{f} = \hat{H}_\mathrm{i}$; which can be understood as the $\kappa\to\infty$ or $f\to 0$ limit
of $\hat{H}_\mathrm{i}$) in Fig.~\ref{fig:GS1.2sr}.
We observe that the condensate fraction is larger for larger $|V|$ and smaller $\kappa$ as expected, while for the short-range case,
the calculated condensate fraction is peaked around $V = -1.96$ before it quickly plunges as $V = -2$ is approached.
We note that the condensate fraction is finite in the short-range case only because the system size is finite, and we expect it to remain finite
in the $L\to\infty$ limit for $\kappa \lesssim 2$ as have been discussed in the main text, though the actual value would be smaller for larger systems
due to the reduction of the finite-size effect.

\begin{figure}[h]
\includegraphics{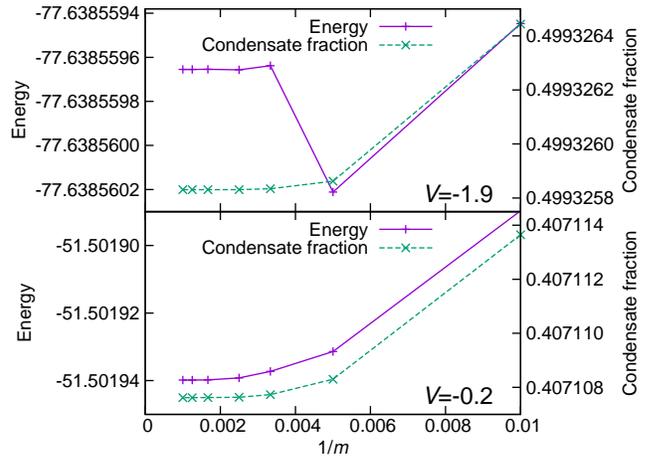}
\caption{(Color online)
Top: The energy of the initial state evaluated by the post-quench Hamiltonian, $\langle \Psi(t=0) \vert \hat{H}_\mathrm{f} \vert \Psi(t=0) \rangle$ (left axis),
and the calculated condensate fraction as functions of the number of states $M$ kept in the initial DMRG calculation
for $L=80$, $N=40$, $f=0.1$, $\kappa=1$, $V=-1.9$.
Bottom: The same for $V=-0.2$.
Note that the calculated energy is not necessarily a decreasing function of $M$ even while DMRG is a variational method,
because the accuracy depends both on the accuracy of $\vert \Psi(t=0) \rangle$ and the representation of $\hat{H}_\mathrm{f}$ obtained in the DMRG calculation.
}
\label{fig:convInit}
\end{figure}

\subsection{Convergence of the initial state}
In Fig.~\ref{fig:convInit} we plot the calculated initial condensate fraction $q(t=0)$ and the energy $\hat{H}_\mathrm{f}$, $\langle \Psi(t=0) \vert \hat{H}_\mathrm{f} \vert \Psi(t=0) \rangle$
as functions of the number of states preserved in each step of the initial finite-system DMRG calculation, for $f=0.1$, $\kappa=1$ and $V=-1.9, -0.2$.
We expect greater numerical difficulty for smaller $\kappa$, because distant pairs of sites are more strongly entangled due to the power-law hopping term in $\hat{H}_\mathrm{i}$,
which hinders convergence in DMRG. The results indicate that the results are well converged for $M \gtrsim 300$.
The slower convergence for smaller $|V| = 0.2$, we believe, is due to the larger entanglement between distant locations on the system due to the larger speed of sound $v$,
and we expect better convergence for $|V| > 0.2$.

\begin{figure}[]
\includegraphics{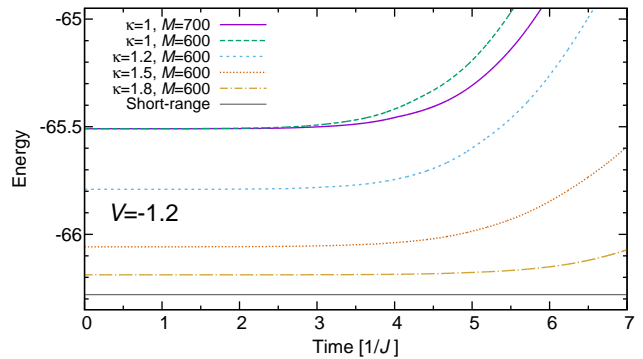}
\caption{(Color online)
The calculated energy $E(t)$ as a function of time for $L=80$, $N=40$, $f=0.1$ and $V = -1.2$, for various values of $(\kappa, M)$ and the short-range limit.
}
\label{fig:EnergyV-12}
\end{figure}

\subsection{Energy conservation for $V=-1.2$}
The energy $E(t) \equiv \langle \Psi(t) \vert \hat{H}_\mathrm{f} \vert \Psi(t) \rangle$ should be a constant of the time $t$
because the wavefunction undergoes a unitary time evolution after the quench, $\vert \Psi(t) \rangle = \exp(i \hat{H}_\mathrm{f} t) \vert \Psi(t=0) \rangle$,
however in the numerical simulation, as the error accumulates, the energy is not fully conserved.
Note that for any normalized state vector $\vert \Psi(t) \rangle$ the energy $E(t)$ should satisfy $E(t)\geq E_0$, in which $E_0$ is the ground state energy for $\hat{H}_\mathrm{f}$.
In our system, we observe that $E(t)$ starts to rapidly increase after some time, earlier for smaller $\kappa$ and smaller $M$.
We have observed that this time roughly corresponds to the time at which the calculated time-dependent exponent $c(t)$ starts to deviate from that obtained with a larger $M$ (see Fig.~\ref{fig2}),
and for $V = -1.2$ and $300 \lesssim M \lesssim 800$, as observed in Fig.~\ref{fig:EnergyV-12},
it also almost coincides with the time at which $q(t)$ reaches the condensate fraction of the short-range model.

\subsection{The case with a larger value of $|V|$: $V = -1.9$}
\begin{figure}[]
\includegraphics{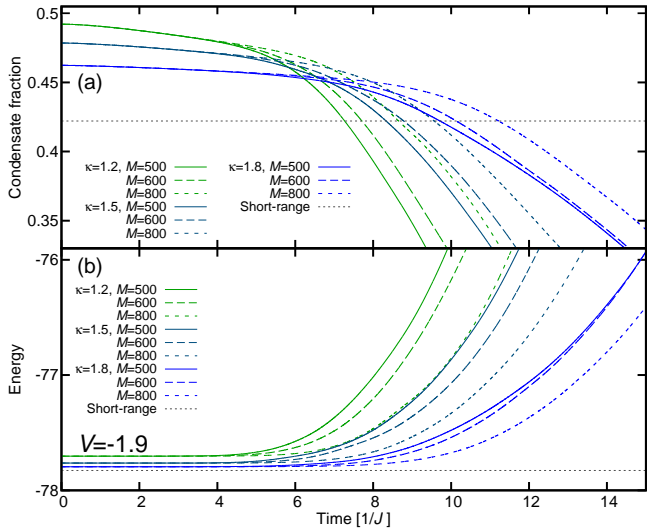}
\caption{(Color online)
Top: The calculated condensate fraction as functions of time, for $\kappa = 1.2, 1.5, 1.8$ and $M=500, 600, 800$ states retained,
for $L=80$, $N=40$, $f=0.1$, $V = -1.9$.
Bottom: The energy, calculated as $\langle \Psi(t) \vert \hat{H}(t>0) \vert \Psi(t) \rangle$, as functions of time for same sets of values of $\kappa$ and $M$.
}
\label{figV19}
\end{figure}

For $V = -1.9$, which is close to the density-wave transition, $v$ is small and the condensate takes a long time to decay.
The results are converged only up to $t\sim 5/J$ for $\kappa = 1.2$, beyond which time the conservation of the energy $E(t)$ is significantly violated.
Between $M = 500$ and $M = 800$, the separation between the plotted values of $q(t)$ in Fig.~\ref{figV19} is significant for $t\gtrsim 6/J$,
and larger for smaller $\kappa$. Correspondingly, $E(t)$ starts to increase around $t = 5/J$. The onset of this increase is not much delayed by increasing $M$,
which suggests that it is numerically challenging to extend the time range of converged simulation for $V \sim -1.9$.

\subsection{The case with a smaller value of $|V|$: $V=-0.8$}
\begin{figure}[]
\includegraphics{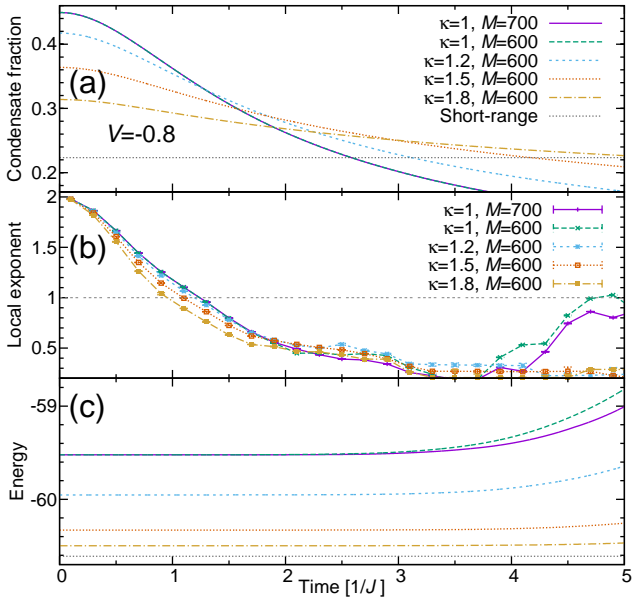}
\caption{(Color online)
a) and b): Same as the corresponding parts in Fig.~\ref{fig2} except that $V=-0.8$ is used here.
c) The energy $E(t) = \langle \Psi(t) \vert \hat{H}_\mathrm{f} \vert \Psi(t) \rangle$, as functions of time for same sets of values of $\kappa$ and $m$.}
\label{figV08}
\end{figure}

As observed in Fig.~\ref{figV08} with $V = -0.8$, for a smaller value of $|V|$ compared to the one in the main text of this manuscript, $V=-1.2$,
the dynamics is faster in time,
However the error accumulates faster for a faster dynamics.
Significant increase in $E(t)$ and deviation between the values of $q(t)$ calculated with $M=600$ and $M=700$ are already visible around $t = 3/J$.

In summary, while the convergence of our t-DMRG simulations are limited in the time range for larger $|V|$,
all results obtained consistently indicate the crossover from the initial Gaussian-like decay of the condensate
to a subsequent stretched exponential decay which is faster for smaller values of $\kappa$.

\end{document}